\documentclass{article}
\usepackage{spconf,amsmath,graphicx}

\usepackage{enumitem}
\usepackage{subcaption}
\usepackage{cite}
\usepackage{amsmath,amssymb,amsfonts}
\usepackage{algorithmic}
\usepackage{graphicx}
\usepackage{hyperref}
\usepackage{textcomp}
\usepackage{xcolor}
\usepackage{booktabs}
\usepackage{multirow}
\usepackage{rotating}
\usepackage{threeparttable}
\usepackage{url}
\setlist{nosep, leftmargin=14pt}
\usepackage{mwe} 

\title{A Deep Ensemble Learning Approach to Lung CT Segmentation for COVID-19 Severity Assessment }
\name{Tal Ben-Haim$^{1}$*, Ron Moshe Sofer$^{1}$*, Gal Ben-Arie$^{2,3}$, Ilan Shelef $^{2,3}$ and Tammy Riklin Raviv$^{1}$}
\address{$^{1}$School of Electrical and Computer Engineering, Ben-Gurion University \\
    $^{2}$Department of Radiology, Soroka Medical Center\\
    $^{3}$Department of Health Sciences, Ben-Gurion University}

\begin{document}
%
\maketitle
\def\thefootnote{*}\footnotetext{Equal contribution}\def\thefootnote{\arabic{footnote}}
\begin{abstract}
We present a novel deep learning approach to categorical segmentation of lung CTs of COVID-19 patients. Specifically, we partition the scans into healthy lung tissues, non-lung regions, and two different, yet visually similar, pathological lung tissues, namely, \textit{ground-glass opacity} and \textit{consolidation}. 
This is accomplished via a unique, end-to-end hierarchical network architecture and ensemble learning, which contribute to the segmentation and provide a measure for segmentation uncertainty.

The proposed framework achieves competitive results and outstanding generalization capabilities for three COVID-19 datasets. Our method is ranked second in a public Kaggle competition for COVID-19 CT images segmentation. 
Moreover, segmentation uncertainty regions are shown to correspond to the disagreements between the manual annotations of two different radiologists. Finally, preliminary promising correspondence results are shown for our private dataset when comparing the patients' COVID-19 severity scores (based on clinical measures), and the segmented lung pathologies. Code and data are available at our repository\footnote{For repository refer to: \scriptsize{\url{https://github.com/talbenha/covid-seg}}}. 


\end{abstract}
\vspace{-0.1cm}
\vspace{-0.4cm}
\section{Introduction}
\label{sec:intro}
\vspace{-0.2cm}

\begin{figure}[t]
\hspace{0cm}{\includegraphics[scale=0.25, trim={0cm 0cm 0cm 0cm},clip ]{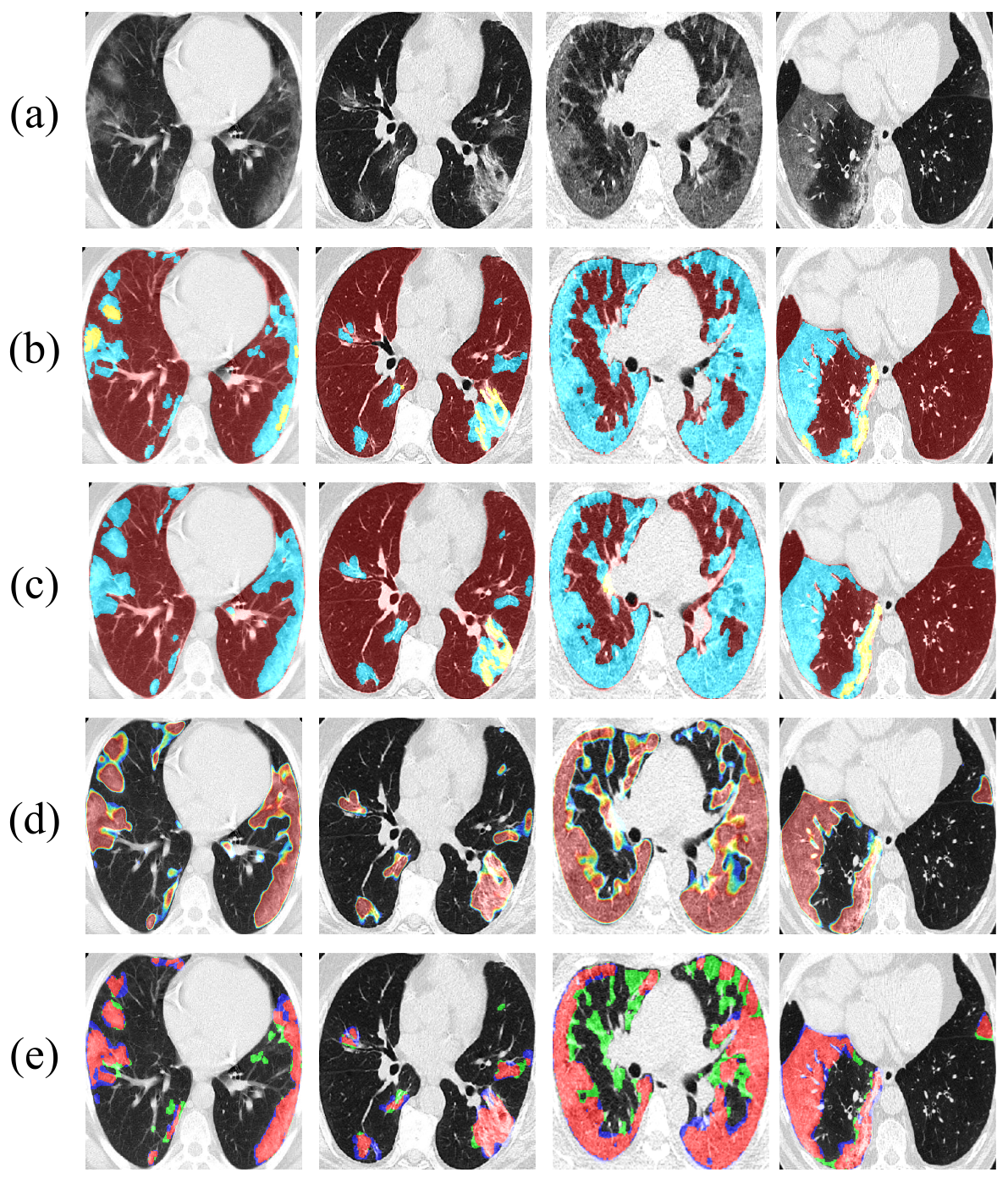}}
\caption{Results for the Radiopaedia Data. 
(a) CT image (b) Rater~1 annotations (c) HU-Net Ensemble prediction. GGO, CON, and healthy lungs are colored in cyan, yellow and red, respectively.
(d) Pathology segmentation uncertainty (jet color map, red is high, blue is low) (e) Raters' consensus. Rater~1, rater~2, and consensus segmentation are colored in green, blue, and red, respectively.
}
\label{fig:Visual_Results}
\vspace{-.3cm}
\end{figure}

Corona virus disease 2019 (COVID-19) is an infectious disease caused by acute respiratory syndrome coronavirus 2 (SARS-CoV-2). The lung disease is characterized by two types of pathology, namely, ground-glass opacity (GGO) and pulmonary consolidation (CON) which can be detected in chest CT scans~\cite{wong2020frequency}. The GGO is manifested by intensity enhancement in the infected lung regions while moderately preserving the bronchial and vascular markings. The CON, which is considered graver, is a region of compressible lung tissue filled with liquid instead of air. It appears brighter and more opaque than the GGO. The extent of each pathology as detected from CT scans can be helpful to assess the expected progress and the severity of the disease~\cite{liu2020ct, sun2020ct,colombi2020qualitative}. However, distinguishing between GGO and CON is difficult even for experienced radiologists due to the gradual change in the level of opacity and the absence of apparent contrast. Moreover, the appearance of each pathological tissue differs between individuals. These challenges are demonstrated in the first two rows of Fig.~\ref{fig:Visual_Results} which present pathological lung CT scans along with GGO and CON segmentations.  
Given the relevance and importance of the problem, many recent studies aim to address lung CT segmentation by estimating the pathological regions. Most of the studies present binary segmentation methods for dividing lung scans into healthy and non-healthy tissues~\cite{chassagnon2021ai,fan2020inf,gao2020dual,sagie2021covid,wu2020jcs,wang2020noise} while only a few present multi-class segmentation techniques for separating between GGO and CON regions~\cite{fan2020inf}.

The proposed framework can distinguish between CON and GGO and has high generalization capabilities thanks to unique hierarchical network architecture and ensemble learning. Utilizing statistics of the ensemble independent predictions allows us to enhance the overall segmentation accuracy and to derive a measure of segmentation uncertainty. 
We evaluated our method using two public datasets, the MedSeg and the Radiopaedia datasets~\cite{jun2020covid}, and a private dataset acquired at the Soroka University Medical Center (SUMC). An expert radiologist in our group provided additional manual segmentations to the annotated public data as well as to slices of each of the SUMC volumes.

Our method is shown to outperform recently published algorithms and is ranked the \textbf{second} in a public Kaggle challenge for COVID-19 CT images segmentation.  Its excellent ability to generalize is demonstrated in comparison to the state-of-the-art nnU-Net~\cite{isensee2021nnu} framework when training on one dataset and testing on another. 
Segmentation uncertainty is validated by showing high correspondence between regions of low prediction confidence and regions of disagreement in manual annotations of the two raters.  
Finally, preliminary promising correspondence results are shown for the SUMC dataset  when comparing the patients' COVID-19 severity scores, calculated based on demographic and clinical measures, and the extent and gravity of the segmented lung pathologies.

\begin{figure*}[t]
\centerline{\includegraphics[scale=0.68,trim={0 0.2cm 0 0cm},clip]{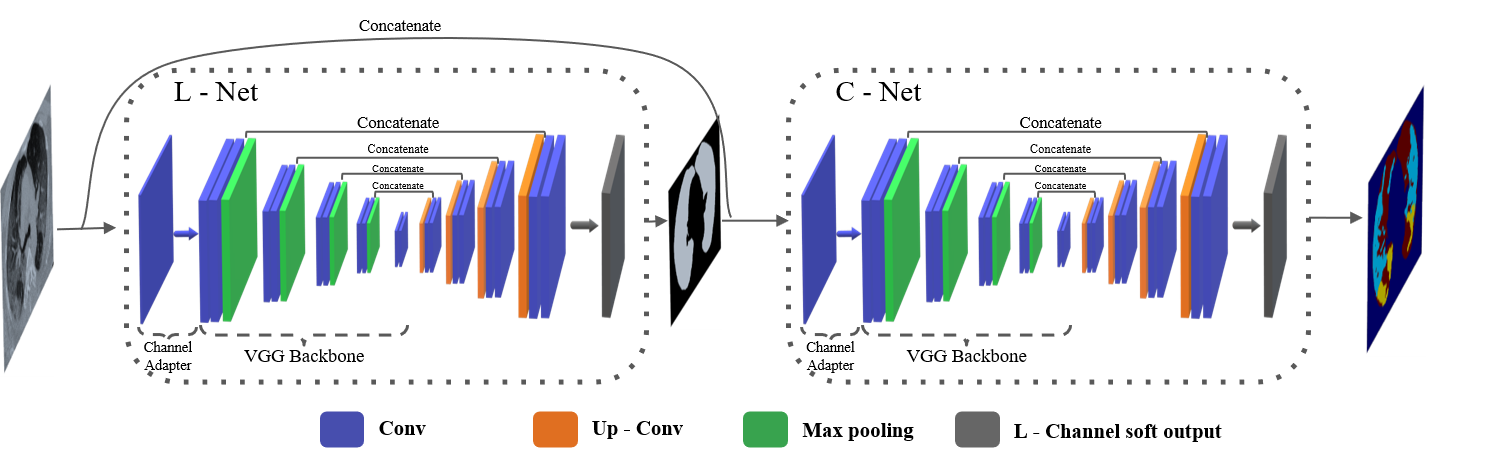}}
\vspace{-2mm}
\caption{The compound hierarchical segmentation architecture. The left-most U-Net segments the lungs' cavity and the right one partitions the lung area into GGO, CON, and healthy tissue.}
\label{fig:Full_Model}
\end{figure*}

\vspace{-0.4cm}
\section{Methods}
\label{sec:methods}
\vspace{-0.2cm}
\textbf{Architecture and loss functions.} The proposed lung CT segmentation framework is based on a cascade of two U-Net networks as illustrated in Fig.~\ref{fig:Full_Model}. The U-Net in is a CNN with an autoencoder structure that is equipped with skip connections between corresponding encoder and decoder layers~\cite{ronneberger2015u}. The U-Net cascade termed HU-Net is trained to perform segmentation in a hierarchical manner, where the lung cavity is first extracted by the left-most U-Net (L-Net) and then the lung region is partitioned by the right-most U-Net (C-Net) into healthy lung tissues, CON and GGO. 
We use both cross-entropy (CE) and Dice loss functions. 
Let $y^{(p)}$ define the one-hot encoded GT label of a pixel $p$ in a batch. Let $L$ define the number of labels.  Then $y^{(p)}_l=1$ and $y^{(p)}_{l'}=0$ for $l' \neq  l$ where, $l,l' \in \{1,\ldots, L\}.$
The Dice loss is composed of $l$-summands  corresponding to $l$-th labels where each is defined as follows:
\begin{equation}
\begin{aligned}
L^l_{\mbox{\scriptsize{Dice}}}  = 1- \frac{2\sum_{p=1}^N y^{(p)}_l \tilde{y}^{(p)}_l}{\sum_{p=1}^N  \big(y^{(p)}_l+\tilde{y}^{(p)}_l\big)+\epsilon} \label{eq:Loss_Dice}
\end{aligned}
\end{equation}
We define the weighted Dice loss by $L_{\mbox{\scriptsize{Dice}}} = \sum_{l=1}^Lw_l L^l_{\mbox{\scriptsize{Dice}}}$.
The weighted categorical cross-entropy (WCE) loss for $N$ pixels in a batch is defined as follows:
\begin{equation}
\begin{aligned}
L_{\mbox{\scriptsize{WCE}}}  = -\sum_{p=1}^N\sum_{l=1}^L w_l y^{(p)}_l \log (\tilde{y}^{(p)}_l)
\label{eq:Loss_WCE}
\end{aligned}
\end{equation}
where $\tilde{y}^{(p)}_l\in[0,1]$ is the network's predicted probability that $l$ is the label of $p.$
To compensate for the imbalance in categories the loss is weighted. We denote by  $w_l$ the weight of the $l$-th term in the loss function. 
The L-Net loss is binary cross-entropy setting $L=2$ and $w_1=w_2=1$.
The C-Net loss is a weighted sum of the WCE, setting $L=4$, and the Dice loss.
\begin{equation}
\begin{aligned}
L_{\mbox{C-Net}}  = \lambda  L_{\mbox\scriptsize{WCE}}+(1-\lambda)L_{\mbox\scriptsize{Dice}}
\label{eq:Loss_Total}
\end{aligned}
\end{equation}
where $\lambda$ is a weight coefficient between the terms.

In addition to the CT scans input the C-Net also receives as input the soft segmentation predictions $\tilde{y}$ 
provided by the L-Net. This allows the application of the proposed network compound in an end-to-end manner such that the C-Net loss backpropagates also to the L-Net thus enhances its training. 
To compensate for the limited training data, the input to the networks are randomly cropped patches of the chest CT scans. We also use transfer learning where the encoder parts of L-Net and C-Net are based on VGG16, which was pre-trained on 3-channel RGB data of the ImageNet~\cite{Deng2009ImageNetAL}.
To adapt the VGG16 to the CT scan's single gray-level channel, we use a convolutional channel adapter.

\noindent\textbf{Ensemble Learning.} We use ensemble learning and train $K$ network compounds. 
The training processes of the networks differ due to the random weight initialization of the decoders' weights as well as the random data augmentation and shuffling. 
Averaging the soft segmentation predictions obtained by each of the $K$ models prior to the \textit{argmax} operation, one can obtain a soft segmentation map, which models segmentation uncertainty. This soft segmentation map quantifies the level of consensus between the predictions of the different networks.

\vspace{-.3cm}
\section{Experiments and Results}\label{sec:exp}
\vspace{-0.2cm}
We evaluate our approach using three lung CT datasets of COVID-19 patients, two public (MedSeg and Radiopaedia~\cite{jun2020covid})and one private (SUMC). We compare our results for the MedSeg dataset to those published in~\cite{fan2020inf}. Our scores for both public datasets were compared to the nnU-Net~\cite{isensee2021nnu}. 
We also contribute additional manual annotations by a radiologist for both datasets. 
We denote by rater~1 and rater~2 the manual annotations provided by the MedSeg radiologist and our colleague, respectively.
We then show that the segmentation uncertainty measure calculated based on the independent predictions provided by the ensemble networks corresponds to the raters' consensus and disagreement. 
Furthermore, we exploit the SUMC data to evaluate the correspondence between patients' clinical risk and the extent and gravity of the COVID-19 pathologies as calculated from the segmentation. The code, additional annotations and annotated SUMC data are available in our \href{https://github.com/talbenha/covid-seg}{repository}.
\vspace{-0.2cm}
\subsection{Datasets}
\vspace{-0.1cm}
The MedSeg dataset consists of 100 chest CT slices from different COVID-19 patients with $512\times 512$  resolution. Radiopaedia dataset consists of chest CT volumes of $8$ different COVID-19 patients. To avoid duplication we used a single 3D scan of each patient with a total number of $790$ slices. 
Annotations include left and right lung cavity lobes, GGO, and CON.
The SUMC dataset contains $23$ CT scans of different patients along with the respective clinical risk scores calculated based on demographic (e.g., age) and clinical (e.g. oxygen saturation) measures.
The training data consists of $74$ CT slices taken exclusively from the MedSeg dataset. 
The test data include the MedSeg slices that were not used for training as well as the entire Radiopaedia and SUMC datasets. 

\begin{table}[t]
\setlength\aboverulesep{0pt}
\setlength\belowrulesep{0pt}
\setlength\tabcolsep{3pt}
\renewcommand{\arraystretch}{1}
  \centering
  \scalebox{0.7}{
    \begin{tabular}{|c|l||c|c|c|c|c|c|}
    \toprule
    \multicolumn{2}{|c||}{\multirow{2}[4]{*}{}} 
    & \multicolumn{2}{c|}{\textbf{GGO}} & \multicolumn{2}{c|}{\textbf{CON}} & \multicolumn{2}{c|}{\textbf{Binary}} \\
\cmidrule{3-8}    \multicolumn{2}{|c||}{} & \textbf{Dice$(\%)$} & \textbf{MHD} & \textbf{Dice$(\%)$} & \textbf{MHD} & \textbf{Dice$(\%)$} & \textbf{MHD} \\

    \midrule
    \midrule
    \multirow{7}[14]{*}{\begin{sideways}\ \ \ \ \ \ \ \ \ \ \textbf{MedSeg}\end{sideways}} & \textbf{InfNet 1} & 65  & -     & 30  & -     & 68  & - \\
\cmidrule{2-8}          & \textbf{InfNet 2} & 62  & -     & 46  & -     & 74  & - \\
\cmidrule{2-8}          & \textbf{U-Net} & 74  & 9.6 $\pm$ 6.6 & 55  & 20.9 $\pm$ 22.5 & 79  & 8.2 $\pm$ 5.4 \\
\cmidrule{2-8}          & \textbf{nnU-Net} & \textbf{79} & 16.2 $\pm$ 23.4 & \textbf{62} & \textbf{17.9 $\pm$ 31.5} & \textbf{83} & 16.7 $\pm$ 23.1 \\
\cmidrule{2-8}          & \textbf{HU - Net} & 74 $\pm$ 1 & 9.9 $\pm$ 6.6 & 53 $\pm$ 5 & 20.4 $\pm$ 21.1 & 78 $\pm$ 1 & 8.9 $\pm$ 5.7 \\
\cmidrule{2-8}          & \textbf{Proposed} & 77  & \textbf{8.4 $\pm$ 6.5} & 57  & 24.0 $\pm$ 28.1 & 81  & \textbf{7.5 $\pm$ 5.8} \\
\cmidrule{2-8}          & \textbf{Rater~2} & 73  & 9.9 $\pm$ 12.7 & 46  & 48.5 $\pm$ 76.2 & 80  & 6.4 $\pm$ 3.2 \\
    \midrule
    \midrule
    \multirow{5}[10]{*}{\begin{sideways}\ \ \ \ \ \ \ \ \textbf{Radiopaedia}\end{sideways}} & \textbf{U-Net} & 71 $\pm$ 4 & 16.5 $\pm$ 18.7 & 59 $\pm$ 4 & 19.2 $\pm$ 33.8 & 72 $\pm$ 3 & 20.0 $\pm$ 17.8 \\
\cmidrule{2-8}          & \textbf{nnU-Net} & 64  & 17.7 $\pm$ 19.9 & 26  & 17.9 $\pm$ 27.0 & 54  & 17.1 $\pm$ 20.5 \\
\cmidrule{2-8}          & \textbf{HU - Net} & 75 $\pm$ 1 & 15.5 $\pm$ 18.4 & 60 $\pm$ 3 & 18.2 $\pm$ 29.1 & 76 $\pm$ 1 & 17.4 $\pm$ 17.1 \\
\cmidrule{2-8}          & \textbf{Proposed} & \textbf{79} & \textbf{11.5 $\pm$ 13.6} & \textbf{68} & \textbf{14.1 $\pm$ 26.8} & \textbf{80} & \textbf{12.3 $\pm$ 13.0} \\
\cmidrule{2-8}          & \textbf{Rater~2} & 78  & 7.8 $\pm$ 10.2 & 74  & 9.1 $\pm$ 22.0 & 82  & 5.0 $\pm$ 4.4 \\
    \bottomrule
    \end{tabular}%
    }
  %
  \vspace{-.2cm}
  \caption{Quantitative COVID-19 Segmentation Results}\label{tab:Quantit_Res}
  \vspace{-.1cm}
\end{table}%
\vspace{-.1cm}
\subsection{Segmentation results and ablation study}
\label{ssec:seg}
\vspace{-.1cm}
{\bf Quantitative results:}
Table~\ref{tab:Quantit_Res} presents quantitative segmentation results using the Dice~\cite{dice1945measures} and the Modified Hausdorff Distance (MHD)~\cite{dubuisson1994modified} scores, obtained for binary (pathology vs. healthy tissues) and multi-class (GGO and CON) for both MedSeg (upper part of the table) and Radiopaedia (lower part of the table) test datasets. 
InfNet~1 and InfNet~2 stand for the results published in~\cite{fan2020inf}, and their variants for multi-class and binary segmentations. We should note that the networks in~\cite{fan2020inf} were trained using scans from both the MedSeg and the Radiopaedia datasets, while all other models were trained only on 74 scans of the MedSeg data.
 The U-Net stands for segmentation performed with the proposed C-Net (without segmentation hierarchy). The nnU-Net, which is also based on the U-Net architecture, was trained according to the training paradigm proposed in~\cite{isensee2021nnu}. 
The HU-Net stands for a single hierarchical model (combined L-Net and C-Net). The results shown are obtained by averaging the segmentation scores of each of the $K=6$ HU-Net models along with the respective standard deviation. 
All of these methods are compared to the proposed HU-Net ensemble - constructed by combining $K=6$ hierarchical models as detailed in Section~\ref{sec:methods}. Rater~2 refers to the evaluation of the manual segmentations of rater~2 with respect to rater~1. The results show that the proposed framework outperforms the InfNet models for all measures for both binary and multi-class segmentations. The comparisons to the U-Net (C-Net segmentation alone) and to a single hierarchical model compose the ablation study demonstrating the advantage of network hierarchy and the strength of the ensemble. The nnU-Net automatically adapts its hyperparameters to the training data. Therefore, it achieves better scores in most of the comparisons when both the training and the test are based on MedSeg dataset (upper part of the table). However, when the networks are trained on MedSeg and tested on Radiopaedia (lower part of the table) the advantage of our method for all measures and tissues is prominent. Specifically, the generalization ability of our model is much better than the nnU-Net's. 
Finally, note that the measures of overlap between the manual annotations of the two raters (quantified by the Dice and MHD scores) are similar to the overlap measures obtained by comparing the proposed automatic segmentation with respect to rater~1 for the binary and multi-class segmentation.  

\begin{figure}[t]
    \centering
    \begin{subfigure}[t]{0.22\textwidth}
        \centering
        \centerline{\includegraphics[scale=0.23, trim={0.00cm 0.00cm 0.00cm 0.00cm},clip ]{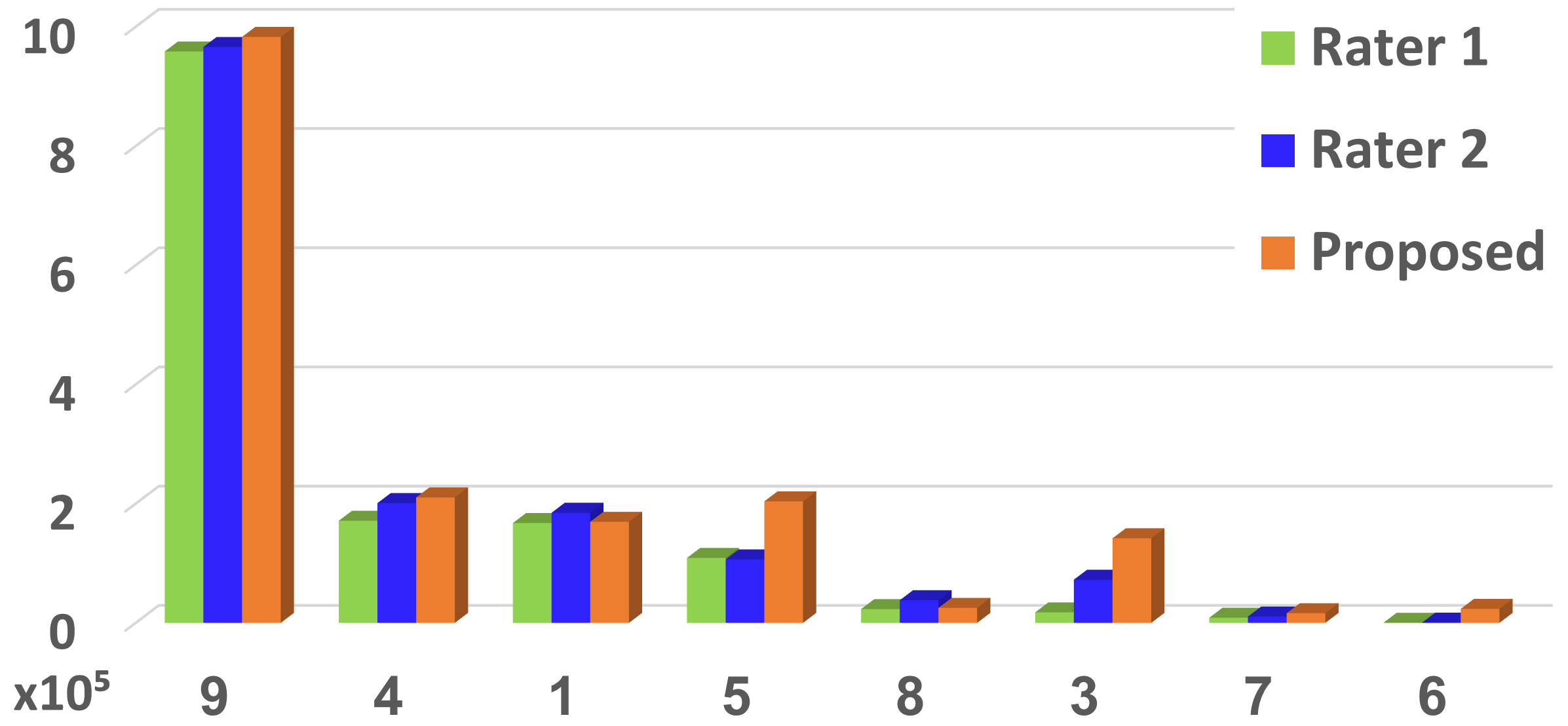}}
        \caption{GGO}
    \end{subfigure}%
    ~
    \begin{subfigure}[t]{0.22\textwidth}
        \centering
        \centerline{\includegraphics[scale=0.23, trim={0.00cm 0.00cm 0.00cm 0.00cm},clip ]{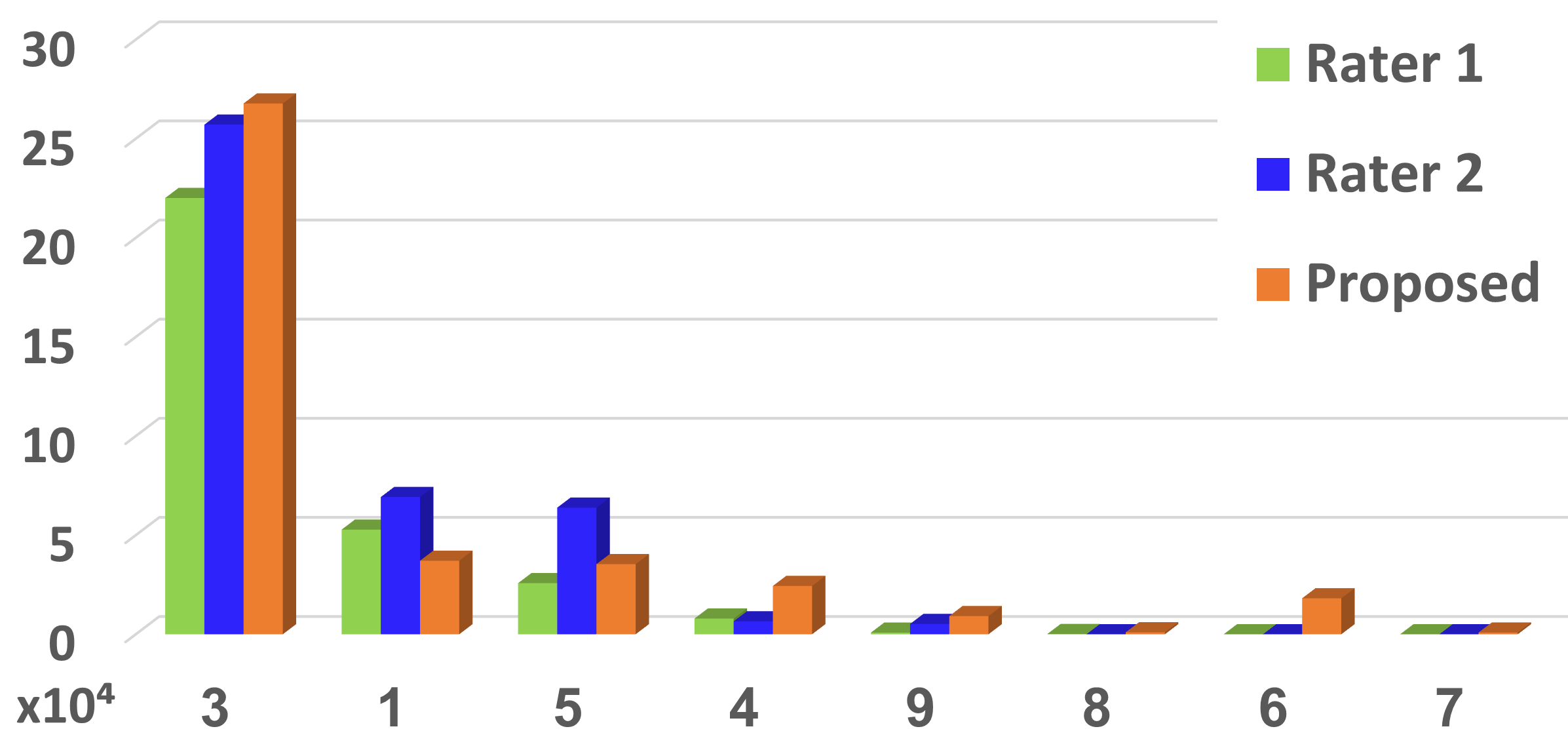}}
        \caption{CON}
    \end{subfigure}
    \vspace{-3mm}
    \caption{Bar plots representing GGO (a) and CON (b) segmented voxels counts for the test eight Radiopaedia volumes. Green, blue and orange colors indicate calculations based on rater~1, rater~2 manual annotations, and the proposed network prediction, respectively.}
\label{fig:Plot_precentage}
\vspace{-3mm}
\end{figure}

\noindent{\bf Kaggle competition:}
Our method (called `ICIP submission') was ranked second in a public Kaggle competition for COVID-19 CT segmentation\footnote{\url{https://www.kaggle.com/c/covid-segmentation/leaderboard}}.
The metric used in the challenge is pixel-wise F1-score which is averaged across all GGO and CON pixels of the test images. The percentage difference between our score and the first place is only $0.08\%$.
\\
{\bf Qualitative results:}
The performance of our model when training on the MedSeg dataset and testing on the Radiopaedia dataset is visually illustrated in Fig.~\ref{fig:Visual_Results}.
Fig.~\ref{fig:Visual_Results}a presents the example CT scans while Fig.~\ref{fig:Visual_Results}b,c present GGO (cyan), CON (yellow), and healthy-lungs (red) segmentations, provided by rater~1, and the proposed HU-Net ensemble, respectively. 
Note the significant visual similarity between the manual segmentation and the network predictions. 
Fig.~\ref{fig:Visual_Results}d presents a color-coded soft segmentation using a jet colormap. The uncertainty measures are derived from the level of consensus between the predictions of the networks in the ensemble. 
Fig.~\ref{fig:Visual_Results}e presents a super-position of the manual segmentations of the two radiologists. Rater~1, rater~2, and consensus segmentation are colored in green, blue, and red, respectively. Comparison of Fig.~\ref{fig:Visual_Results}d and e shows that segmented regions with high uncertainty (colored in green and yellow) mostly correspond to regions for which the radiologists' annotations do not overlap, while regions of consensus (marked in red) either between networks or raters are similar. Finally, Fig.~\ref{fig:Plot_precentage} presents bar plots of the number of voxels classified as GGO and CON for each of the eight Radiopaedia volumes that were tested. Note the similarity in volume estimates between our method and the manual segmentations of both raters.


\begin{figure}[t]
    \centering
    \begin{subfigure}[t]{0.23\textwidth}
        \centering
        \centerline{\includegraphics[scale=0.24, trim={0.00cm 0.20cm 0.00cm 0.20cm},clip ]{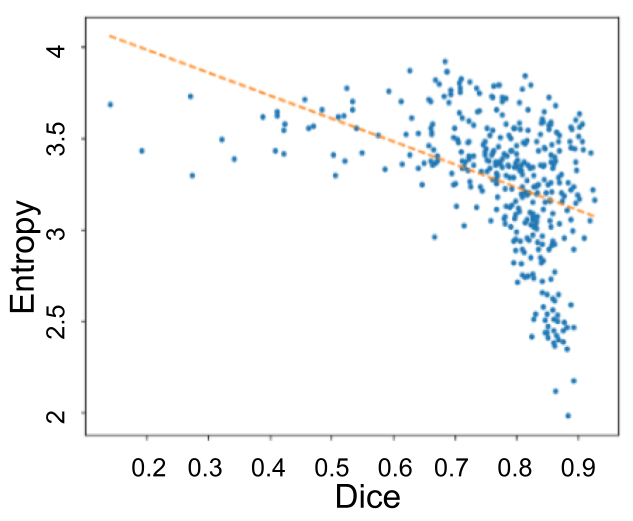}}
        \caption{Uncertainty}\label{fig:scatter_uncertainty}
    \end{subfigure}%
    ~
    \begin{subfigure}[t]{0.23\textwidth}
        \centering
        \centerline{\includegraphics[scale=0.21, trim={.5cm .8cm .80cm 0.00cm},clip ]{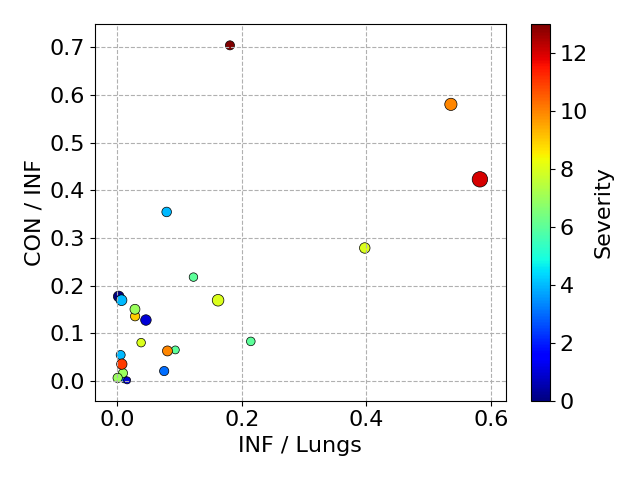}}
        \caption{Severity}\label{fig:severity_scatter}
    \end{subfigure}
    \vspace{-3mm}
    \caption{\textbf{Scatter plots.} (a) Each circle represents a single slice of the test datasets. x-axis: Rater annotation overlaps, calculated using the Dice scores. y-axis: The entropy of the soft segmentation maps obtained from the ensemble predictions. (b) Each circle represents measurements of a single patient. x-axis: The estimated ratio between infection and the entire lungs cavity. y-axis: the estimated ratio between CON and the infection regions (CON+GGO). The patients' clinical risk is color-coded (jet). The radii of the circles is correlated with the mean standard deviation of the x,y axes according to the different models in the ensemble.}
\label{fig:Plot_precentage}
    \vspace{-2mm}
\end{figure}

\noindent\textbf{Segmentation uncertainty:}
Having $K$ segmentation predictions from the network ensemble allows us to both enhance performance and to provide a soft segmentation map that quantifies segmentation uncertainty. Fig.~\ref{fig:Visual_Results}d,e visually asses the correspondence between raters' disagreement and high uncertainty regions. To quantify segmentation uncertainty, we compute the entropy of the value distribution of the soft segmentation map where higher entropy indicates higher uncertainty. The scatter plot, presented in Fig.~\ref{fig:scatter_uncertainty} shows that the Dice measure between the segmentations of the two raters, which indicate their correspondences, negatively correlates with the entropy (Pearson $-0.422$, p-value $1.7e-17$). 
\vspace{-0.3cm}
\subsection{Severity Assessment}

To test the hypothesis that patient's clinical risk is influenced by the extent and gravity of the infectious regions we estimated the 
lung cavity, CON, and GGO volumes of the SUMC patients (based on their segmentations) and calculated their ratios.
Segmentation accuracy was calculated via the available GT annotations (a single annotated slice of each volume) receiving Dice scores of 0.55, 0.54, and 0.655 for GGO, CON, and the entire infectious regions, respectively. Fig.\ref{fig:severity_scatter} presents a scatter plot, where each circle represents measurements of a single patient. The scatter plot illustrates the correspondence between the infection extent (measured by the ratio between the infection regions and the lung cavity) and its gravity (measured by the ratio between the CON regions and the infectious regions).
Having the different ensemble predictions we could calculate the average standard deviations of the volume ratios (x- and y- axes).


\vspace{-0.2cm}
\section{Conclusions}
\vspace{-0.3cm}
We presented a novel deep learning approach for multi-class segmentation of healthy and pathological tissues in lung CT scans. While the method is demonstrated for imaging data of COVID-19 patients - it can be easily adapted to different multi-class segmentation problems. Constructing an ensemble of deep hierarchical models allows us to perform high-quality segmentation by accurately separating between visually similar tissues. Moreover, we tested segmentation uncertainty - comparing the consensus of the obtained ensemble results to the consensus between two human raters. Finally, we demonstrated how lung pathology quantification could provide valuable information to support the evaluation of disease severity and clinical risk.   

\section{ACKNOWLEDGEMENTS}
This study was partially supported by the Israel Ministry of Science, Technology and Space (MOST 3-16902 T.R.R.).



\bibliographystyle{IEEEbib}
\bibliography{strings,refs}

\end{document}